%% file: CKM2010-CLEO.tex
\documentclass[12pt]{article}
\usepackage{graphicx}
\RequirePackage{xspace}

\def\to                 {\ensuremath{\rightarrow}\xspace}

\def\piz   {\ensuremath{\pi^0}\xspace}
\def\pip   {\ensuremath{\pi^+}\xspace}
\def\pim   {\ensuremath{\pi^-}\xspace}

\def\Kbar  {\kern 0.2em\overline{\kern -0.2em K}{}\xspace}

\def\Kz    {\ensuremath{K^0}\xspace}
\def\Kzb   {\ensuremath{\Kbar^0}\xspace}
\def\KzKzb {\ensuremath{\Kz \kern -0.16em \Kzb}\xspace}
\def\Kp    {\ensuremath{K^+}\xspace}
\def\Km    {\ensuremath{K^-}\xspace}
\def\Kpm   {\ensuremath{K^\pm}\xspace}

\def\KpKm  {\ensuremath{\Kp \kern -0.16em \Km}\xspace}
\def\KS    {\ensuremath{K^0_{\scriptscriptstyle S}}\xspace}
\def\KL    {\ensuremath{K^0_{\scriptscriptstyle L}}\xspace}
\def\Kbar  {\kern 0.2em\overline{\kern -0.2em K}{}\xspace}

\def\Dbar    {\kern 0.2em\overline{\kern -0.2em D}{}\xspace}

\def\Dz      {\ensuremath{D^0}\xspace}

\def\Dzb     {\ensuremath{\Dbar^0}\xspace}
\def\DzDzb   {\ensuremath{\Dz {\kern -0.16em \Dzb}}\xspace}
\def\Dp      {\ensuremath{D^+}\xspace}
\def\Dm      {\ensuremath{D^-}\xspace}

\def\DpDm    {\ensuremath{\Dp {\kern -0.16em \Dm}}\xspace}

\def\Bbar    {\kern 0.18em\overline{\kern -0.18em B}{}\xspace}

\def\Bub     {\ensuremath{B^-}\xspace}
\def\Bpm     {\ensuremath{B^\pm}\xspace}

\def\Bm      {\ensuremath{\Bub}\xspace}

\def\dD    {\ensuremath{\Delta\delta_{D}}\xspace}

\newcommand\pubnumber{}
\newcommand\pubdate{\today}

\def\STFC{STFC Rutherford Appleton Laboratory,\\
Chilton, Didcot, Oxfordshire, OX11 0QX, UK}
\def\support{\footnote{On behalf of the CLEO Collaboration.}}

\def\Title#1{\begin{center} {\Large #1 } \end{center}}
\def\Author#1{\begin{center}{ \sc #1} \end{center}}
\def\Address#1{\begin{center}{ \it #1} \end{center}}

\newcommand\pubblock{\rightline{\begin{tabular}{l} \pubnumber\\
         \pubdate  \end{tabular}}}
\newenvironment{Abstract}{\begin{quotation}  }{\end{quotation}}
\newenvironment{Presented}{\begin{quotation} 
      \begin{center}\begin{large}}{\end{large}\end{center} \end{quotation}}

\input econfmacros.tex

\begin{document}
\begin{titlepage}
\pubblock

\vfill
\Title{CLEO-c inputs to the determination of the CKM angle $\gamma$}
\vfill
\Author{ Stefania Ricciardi\support}
\Address{\STFC}
\vfill
\begin{Abstract}
\noindent
The strong-phase differences between $D^0$ and $\bar{D^0}$ decays to common final states are crucial parameters in  the determination of the CKM angle $\gamma$ from $B\to DK$ modes. The first quantum-correlated measurements of these parameters in several $D$ decay modes have been performed with the CLEO-c data at the $\psi(3770)$ resonance. Studies for $D\to K^0_S K^+K^-$, $D\to K^0_S\pi^+\pi^-$, $D\to K^+\pi^-$, $D\to K^+\pi^-\pi^0$, and $D\to K^+\pi^-\pi^+\pi^-$ are reviewed. 
\end{Abstract}
\vfill
\begin{Presented}
\small{ Presented at CKM2010,\\ 
the~6th~International~Workshop on the~CKM~Unitarity~Triangle,\\
University of Warwick, UK, 6-10 September 2010}
\end{Presented}
\vfill
\end{titlepage}
\def\thefootnote{\fnsymbol{footnote}}
\setcounter{footnote}{0}

\section{Introduction}

Quantum-correlated $\Dz\Dzb$ pairs from the coherent $\psi(3770)$ decay provide direct sensitivity to the strong-phase difference, $\Delta\delta_D$, between the $\Dz$ and the $\Dzb$ decay to a common final state. This phase has an important role in the measurement of the CKM angle $\gamma$ from decays of the $B\to DK$ family.\footnote{Here, and in the following, $D$ denotes either \Dz or \Dzb.}

In these proceedings, we describe the recent CLEO-c measurements of \dD for $D\to\KS\Kp\Km$ and $D\to\KS\pip\pim$~\cite{bib:PRD-KShh} and their impact on the determination of $\gamma$. We also mention a new preliminary results for $D\to\Kp\pim$, and we conclude by recalling the results on \dD and the coherence factors for $D\to K^+\pi^-\pi^0$ and $D\to K^+\pi^-\pi^+\pi^-$. All these studies are based on the full CLEO-c dataset at the $\psi(3770)$ resonance, corresponding to 818~pb$^{-1}$ of $e^+e^-$ collisions.

\section{Measurements of \dD for $D\to K^0 h^+h^-$ decays}

The most precise measurements of $\gamma$ to date~\cite{bib:BaBar2010, bib:Belle2010} are based on the differences between
the $D\to \KS h^+h^-$ ($h = K$ or $\pi$) Dalitz plots for $B^-$ and $B^+$ decays to $DK$. These measurements use a model to describe the $D\to \KS h^+h^-$ decay amplitude. The systematic uncertainty on $\gamma$ introduced by the model assumptions is estimated to be between 3$^\circ$ and 9$^\circ$, which is less than the current statistical uncertainty but will be a limiting factor in future measurements. To overcome such limitation, a model-independent method has been developed~\cite{bib:GGSZ, bib:Bondar}. 
The method relies on measurements of the $B^+$ and $B^-$ decay rates in bins of the corresponding $D\to\KS h^+h^-$ Dalitz plots, and on measurements of the amplitude-weighted average cosine ($c_i$) and sine ($s_i$) of \dD over the same bins. 

Measurements of the $c_i$ and $s_i$ parameters are performed by exploiting the quantum-coherence of $D\bar{D}$ pairs from the $\psi(3770)$ decay and a {\em double-tagging} technique.
The clean environment and the excellent performance of the detector allow CLEO-c to reconstruct both the $D$ decay of interest and the recoiling $D$ meson ({\em D-tag}) with high efficiency and purity. 
In particular, the $c_i$ parameters can be determined from the event yields of CP-tagged $D$ decays, while mixed CP-tagged events, such as $D\to\KS\pip\pim$ vs $D\to\KS\pip\pim$, are sensitive to both $c_i$ and $s_i$.
Analogous quantities ($c_i\sp{\prime}$ and $s_i\sp{\prime}$) can be defined for the $D\to K^0_L h^+ h^-$ decay. These primed quantities can exhibit small differences from the corresponding unprimed ones due to additional doubly-Cabibbo-suppressed contributions in the $D\rightarrow K^0_L h^+ h^-$ amplitude. 

The parameters $c_i$, $s_i$, $c_i\sp{\prime}$, and $s_i\sp{\prime}$ are simultaneously extracted, with a maximum likelihood fit, from the background-subtracted and efficiency-corrected bin yields for CP-tagged, $\KS h^+ h^-$-tagged and flavour-tagged $D\to K^0_{S,L}\Kp\Km$ (or $D\to~K^0_{S,L}\pip\pim$) decays. Using decays of the type $\KS hh$ vs $\KL hh$ greatly improves the precision on the $c_i$ and $s_i$ coefficients.

\begin{figure}[htb]
\centering
\includegraphics[height=2.5in]{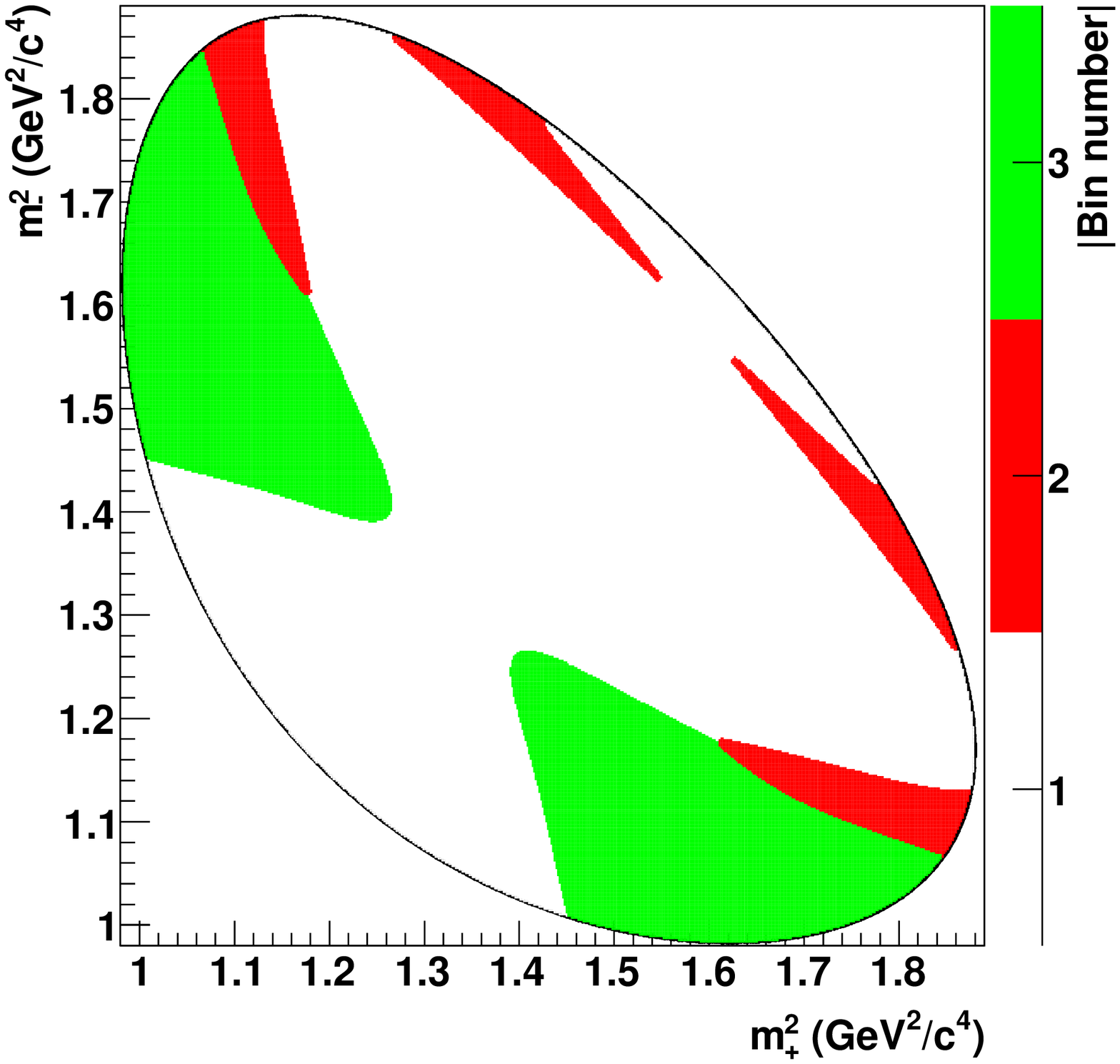}
\includegraphics[height=2.7in]{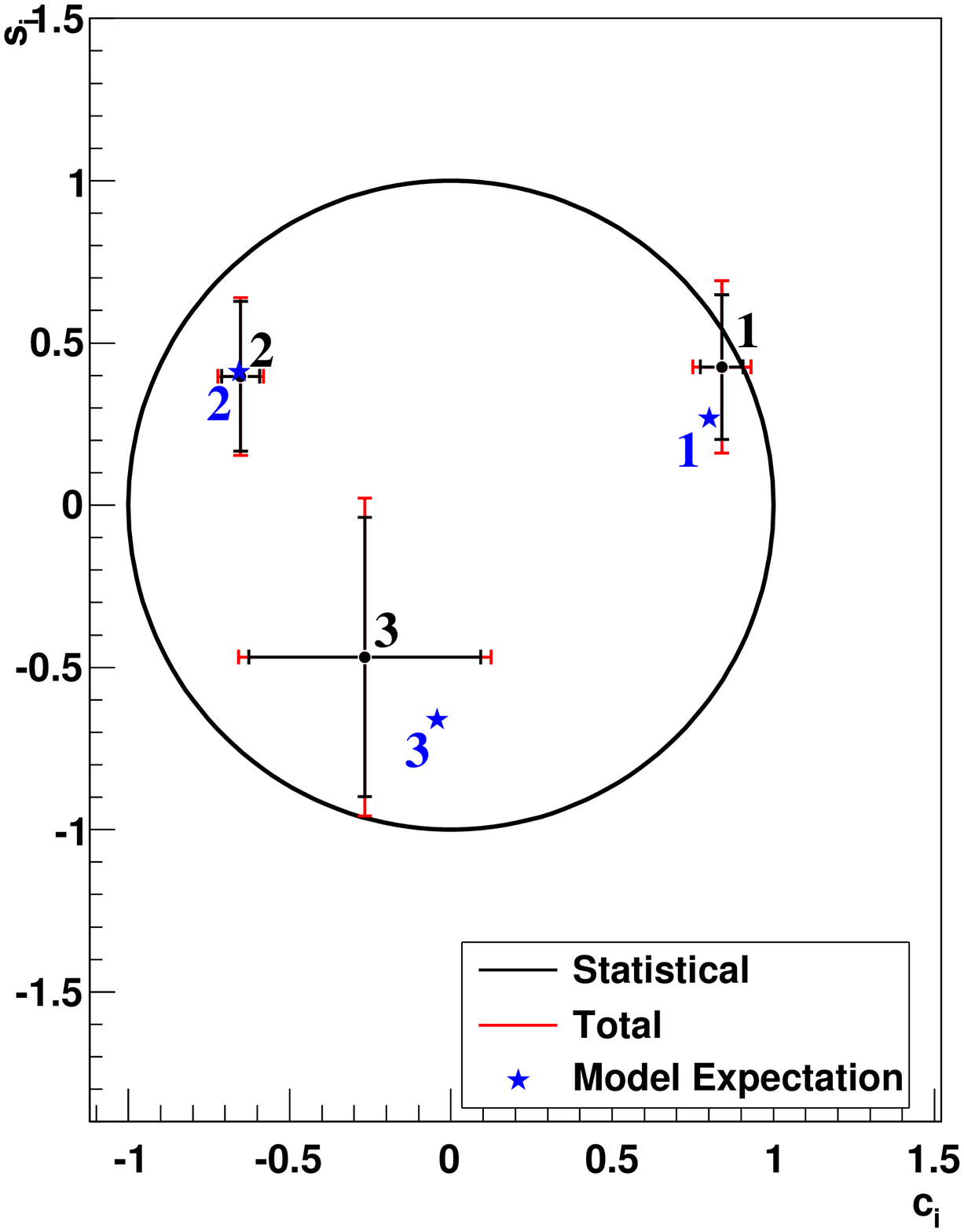}
\caption{$D\to \KS K^+K^-$: an example of $\Delta\delta_D$ binning of the Dalitz plot (left); corresponding results for $c_i$ and $s_i$ (right). Error bars indicate the measured values; stars indicate the predicted values from the BaBar model~\cite{bib:BaBar2010}.}
\label{fig:KSKK}
\end{figure}

\begin{figure}[!hbpt]
\centering
\includegraphics[height=2.5in]{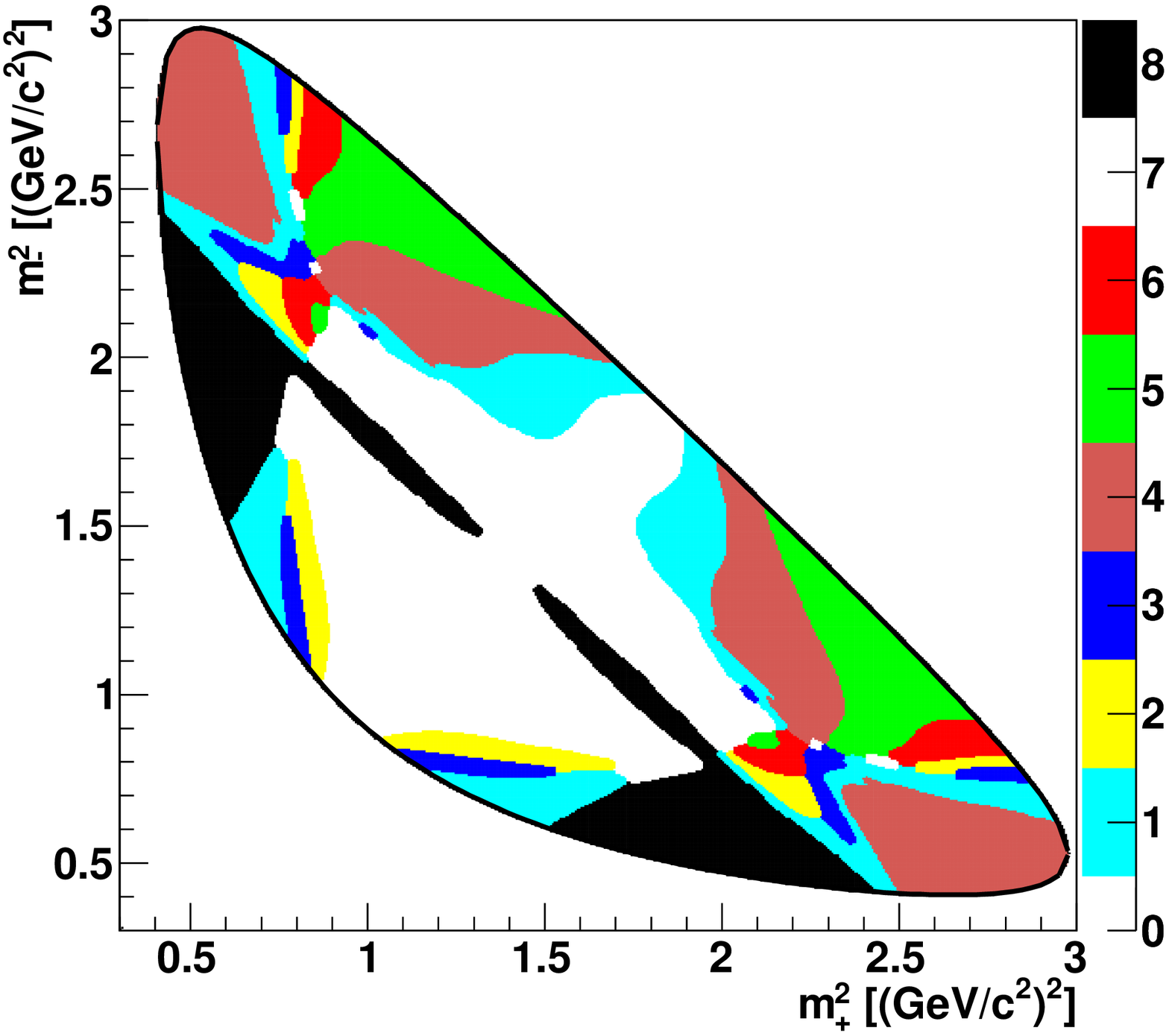}
\includegraphics[height=2.5in]{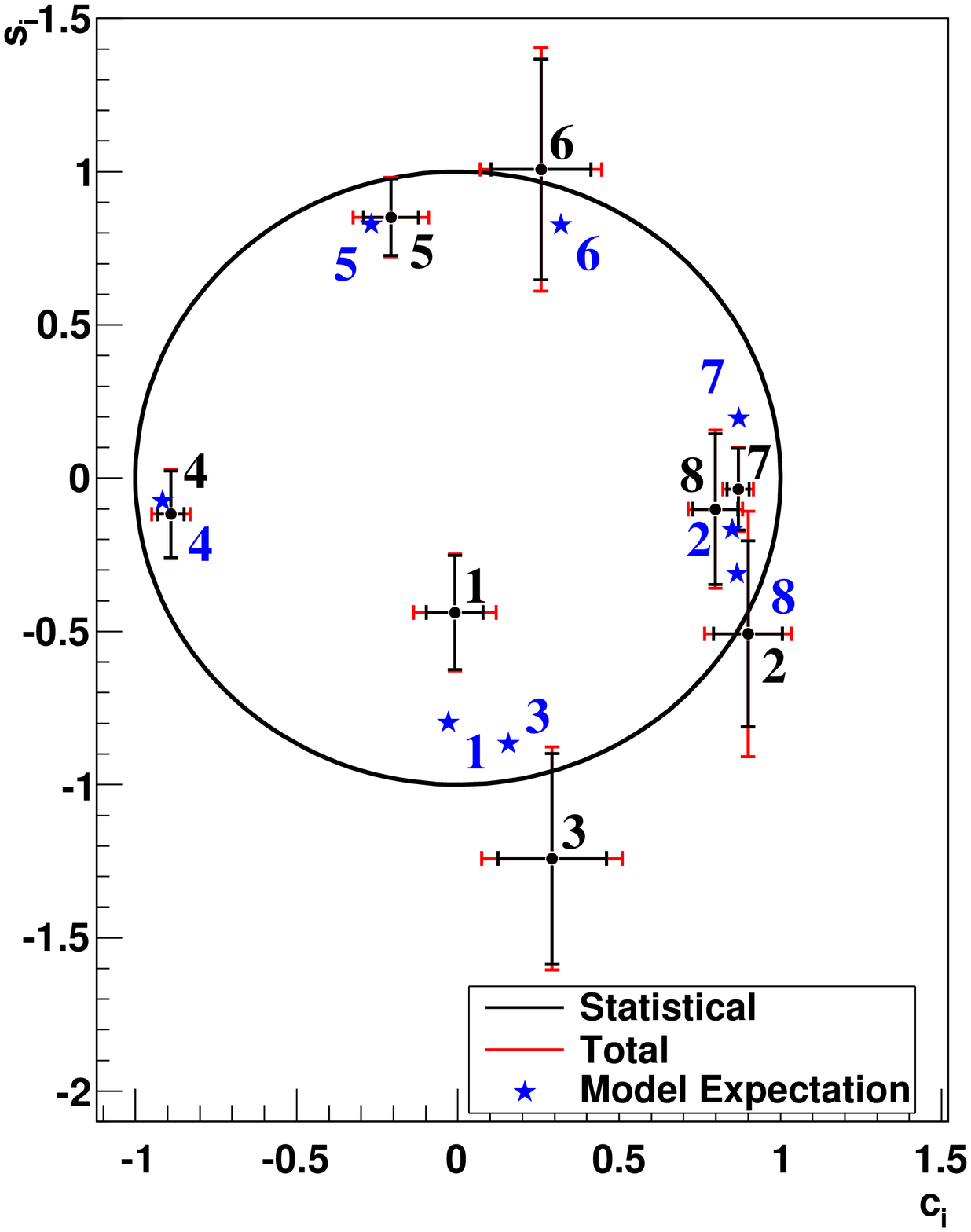}
\caption{$D\to \KS\pip\pim$: an example of $\Delta\delta_D$ binning of the Dalitz plot (left); corresponding results for $c_i$ and $s_i$ (right). Error bars indicate the measured values; stars indicate the predicted values from the BaBar model~\cite{bib:BaBar2008}.}
\label{fig:KSpipi}
\end{figure}

Different number of bins and bin shapes are considered for each decay in order to provide alternatives suitable for the size of different B-decay samples and for different background scenarios. The results for all the considered binning choices are presented in Ref.~\cite{bib:PRD-KShh}. 
The measured values of $c_i$ and $s_i$ for $D\rightarrow\KS K^+K^-$ and  $D\rightarrow\KS\pip\pim$, shown in Fig.~\ref{fig:KSKK} 
and Fig.~\ref{fig:KSpipi} respectively, correspond to bins of equal intervals in $\dD$, according to the BaBar models for $D\to\KS\Kp\Km$~\cite{bib:BaBar2010} and $D\to\KS\pip\pim$~\cite{bib:BaBar2008}. It was shown that this binning choice gives increased sensitivity to $\gamma$ compared to rectangular bins, 
without introducing any model uncertainty~\cite{bib:BondarPoluektov}. 

The measured values are in good agreement with the predicted values, computed from existing models. The systematic uncertainties
are small compared to the statistical uncertainties and mainly of experimental origin, with the largest ones arising from the background subtraction procedure. 

The measurement uncertainties on $c_i$ and $s_i$ will induce a systematic uncertainty on $\gamma$, which has been evaluated to be between 3.2$^\circ$ and 3.9$^\circ$ for $\Bpm\to~D(\KS\Kp\Km)\Kpm$, and between $1.7^{\circ}$ and $3.9^{\circ}$ for $\Bpm\to~D(\KS\pip\pim)\Kpm$, depending on the binning choice. These residual errors, which are due mainly to the limited size of the CLEO-c data sample, will replace the corresponding model uncertainties in future measurements based on the binned approach. 
There is a small price to pay for the model-independence, which is a statistical loss in the sensitivity to $\gamma$ compared to the unbinned method due to the use of discrete information in the binned procedure. This loss has been estimated to be about 10\%.
Discussions on the ultimate precision which can be achieved by future larger samples collected at the $\psi(3770)$ by BES-III can be found elsewhere~\cite{bib:Spradlin, bib:Poluektov} in the proceedings of this conference.

\section{An update on the $D\to K \pi$ study}

Another powerful way to measure $\gamma$~\cite{bib:ADS} exploits the enhanced interference effects in 
$B\to D(K\pi)K$ decays. 
The method is based on the measurement of four separate decay rates from which $\gamma$, and other $B$ and $D$
decay parameters can be determined. Constraints from external measurements of the $D$ decay parameters are crucial, as they allow experiments to reduce the number of degrees of freedom in the extraction of $\gamma$ and improve the sensitivity. 
In particular, $\Delta\delta_D^{K\pi}$, the relative phase between the doubly-Cabibbo-suppressed $\Dz\to\Kp\pim$ and the Cabibbo-favoured $\Dzb\to\Kp\pim$ amplitudes, has been measured by CLEO using 281 pb$^{-1}$ of $\psi(3770)$ data~\cite{bib:CLEO-Kpi}. This phase 
also connects measurements of the $D$-mixing parameters $x,y$ and $y'\equiv y\cos\dD -x\sin\dD$.

A new preliminary result based on the full $\psi(3770)$ data-set is presented here. 

Similarly to other \dD measurements, sensitivity to the
strong-phase is achieved via a double-tagging technique. Several $CP$, mixed-$CP$, and semileptonic $D$-tags are used to extract $\Delta\delta_D^{K\pi}$, $r_D^{K\pi}$(i.e., the magnitude of the ratio between the doubly Cabibbo-suppressed and the favoured $D\to K\pi$ amplitude) and the mixing parameters. Among the improvements in the analysis, in addition to the larger data-sample, we note the inclusion of several $D$-tags not previously considered, such as: $CP$-eigenstates with $\KL$ in the final states; the semi-muonic mode $K\mu\nu$, which doubles the semileptonic yield, from which the $D$-mixing parameter $y$ is determined; the doubly-Cabibbo-suppressed modes vs semileptonic modes; the $\KS\pip\pim$ decay in bins of the Dalitz plot, which provides information on sine of $\Delta\delta_D^{K\pi}$; and, a decay with two missing particles, $\KL\piz$ vs $Ke\nu$.

The large number of $D$-tags allows CLEO-c to extract for the first time the $D$-mixing parameters, $r_D^{K\pi}$ 
, and both the sine and the cosine of \dD  without external constraints. The new preliminary results on the strong-phase are: $\cos\Delta\delta_D^{K\pi} = 0.98^{+0.27}_{-0.20}\pm 0.08$, and $\sin\Delta\delta_D^{K\pi} = -0.04 \pm 0.49 \pm 0.08$. 
A fit with external constraints on the $D$-mixing parameters is also performed and further improves the measurement of \dD. The preliminary result is $\Delta\delta_D^{K\pi}= (15^{+11}_{-7}\pm 7)^{\circ}$.

\section{Measurements of \dD and of the coherence factor for $D\to\Km\pip\pi^0$ and $D\to\Km\pip\pim\pip$}

Additional sensitivity to $\gamma$ can be obtained from $B\to DK$, where $D$ decays to multi-body quasi-flavour-specific final states, such $\Km\pip\piz$ and $\Km\pip\pim\pip$~\cite{bib:ADS, bib:AS}.
For these decays, the variation of \dD over the multi-body phase-space leads to the introduction of the so-called ``coherence factors'', $R_{K\pi\piz}$ and $R_{K3\pi}$, 
which multiply the interference terms sensitive to $\gamma$ in the corresponding $B\to DK$ decay rates.
If, for a given decay channel, there is only a single intermediate resonance or a few non-interfering resonances, the coherence factor will be close to one, and the sensitivity to $\gamma$ will be maximal. If there are many overlapping intermediate resonances the coherence factor will tend towards zero, limiting the sensitivity to $\gamma$. 

The coherence factors and the average strong-phase differences over the multi-body phase-space, $\delta_{K\pi\piz}$ and $\delta_{K3\pi}$, have been measured by CLEO-c~\cite{bib:coherence}, using $D$ decays tagged by either $CP$-eigenstates or $\Km\pip$, $\Km\pip\piz$, and $\Km\pip\pim\pip$, where the tag kaon charge is the same as the signal.  A $\chi^2$ fit to the yields gives: 
$R_{K\pi\piz} = 0.84 \pm 0.07$, $\delta_{K\pi\piz} = (227^{+14}_{-17})^{\circ}$, $R_{K3\pi} = 0.33^{+0.26}_{-0.23}$, and 
$\delta_{K3\pi} = (114^{+26}_{-23})^{\circ}$.

These results show that $D\to\Km\pim\piz$ is highly coherent and indicate that $D\to\Km\pip\pim\pip$ is less so.
However, it has been shown by a study within LHCb~\cite{bib:globalfit} that 
including this decay in a global fit to several $B\to DK$ modes enhances the sensitivity to the magnitude of the amplitude ratio between $\Bm\to\Dz K$ and $\Bm\to \Dzb K$ and leads to a better overall sensitivity to $\gamma$.
The observed improvement in the $\gamma$ precision, when CLEO-c constraints are used for $D\to\Km\pip$ and $D\to\Km\pip\pim\pip$, is roughly equivalent to that which would come from a doubling of the LHCb dataset. 
The mode $D\to \Km\pip\piz$ is not yet included in the LHCb global fit, but it is expected that the LHCb sensitivity to $\gamma$ will further benefit from the inclusion of the CLEO-c constraints on this mode.


\section{Conclusions}

Quantum-correlation effects in the $\psi$(3770) decay have allowed CLEO-c to measure the strong-phase parameters of
$D$ decays to $K^0_S K^+K^-$, $K^0_S\pi^+\pi^-$, $K^+\pi^-$, $K^+\pi^-\pi^0$, and $K^+\pi^-\pi^+\pi^-$. 
These results will have a significant impact in the determination of $\gamma$ from $B\to DK$ decays.
There are other $D$ decay channels of interest to the measurement of $\gamma$, which can be studied with CLEO-c data. Results for $D\to\KS \Kp\pim$ are anticipated to appear soon.


\end{document}

%% file: econfmacros.tex
\def\beq{\begin{equation}}
\def\eeq#1{\label{#1}\end{equation}}
\def\eeqn{\end{equation}}

\def\beqa{\begin{eqnarray}}
\def\eeqa#1{\label{#1}\end{eqnarray}}
\def\eeqan{\end{eqnarray}}

\let\bar=\overbar

\def\Dslash{\not{\hbox{\kern-4pt $D$}}}
\def\dslash{\not{\hbox{\kern-2pt $\del$}}}

\def\msb{{\bar{\ssstyle M \kern -1pt S}}}

